\documentclass[preprintnumbers,amsmath,amssymb,showpacs]{revtex4}
\usepackage{graphicx}% Include figure files
\usepackage{dcolumn}% Align table columns on decimal point
\usepackage{bm}% bold math
\usepackage{booktabs}

\begin{document}

\title{Entangler and analyzer for multiphoton maximally entangled states using weak nonlinearities}

\author{Dong Ding$^{1,2}$}
\author{Fengli Yan$^{1}$}
 \email{flyan@hebtu.edu.cn}

\affiliation {$^1$ College of Physics Science and Information Engineering, Hebei Normal University, Shijiazhuang 050024, China\\
$^2$Department of Basic Curriculum, North China Institute of Science
and Technology, Beijing 101601, China}

\date{\today}

\begin{abstract}

In the regime of weak nonlinearity we present two general  feasible schemes.  One  is an entangler for generating any one of the $n$-photon Greenberger-Horne-Zeilinge (GHZ) states and Bell states. After the interactions with cross-Kerr nonlinear media, a phase gate followed by a measurement on the probe beam, and appropriate local operations via classical feed-forward, one can obtain the desired states in a nearly deterministic way. Another scheme is an analyzer for multiphoton maximally entangled states, which  is taken as a further application of the above entangler. In this scheme, all of the $2^n$ $n$-photon GHZ states can, nearly deterministically, be discriminated. Furthermore, an efficient two-step nondestructive Bell-state analyzer is designed.

\end{abstract}


\maketitle

\section{Introduction}

Quantum entanglement has attracted much attention over the last 20 years, partly because of its potential for some key quantum processes, for example, quantum cryptography \cite{BB84,Ekert91}, quantum dense coding \cite{Bennett and Wiesner 1992}, quantum teleportation \cite{B93}, quantum computation \cite{Raussendorf and Briegel 2001},  and so on. Although most of the current protocols are concerned with bipartite systems, multipartite entanglement has also had potential for applications in quantum information processing, such as the GHZ argument for testing local realism \cite{Klyshko1993}. For genuine tripartite entanglement, it has been shown that there exist at least two inequivalent classes,  the GHZ type and W type \cite{GHZ1990,W2000}. These two different types of entangled states are not equivalent and cannot be converted to each other by local operations and classical communications. The GHZ state of $N$ qubits, $|\text{GHZ}\rangle_{N}=(|0\rangle^{\otimes N}+|1\rangle^{\otimes N})/\sqrt{2}$, is a simple generalization of the three-qubit GHZ state and can be considered as the maximally entangled multi-qubit state \cite{Otfried2009, Yan2011}.

A quantum nondemolition (QND) measurement \cite{Milburn1984, Imoto1985} is of considerable importance in the quantum measurement theory. The QND measurements are designed to perform repeated measurements of quantum states \cite{Braginsky1996} and, in optics, have explored the ultimate quantum limitations to extract (non-destructive) the information encoded in a laser beam \cite{Grangier1998}. With the QND detection based on the optical Kerr effect, recently, Barrett \emph{et al}. \cite{Barrett2005} proposed a scheme to construct a nondestructive Bell-state analyzer from cross-Kerr nonlinearities \cite{Munro2005}. In their scheme, an analyzer has been suggested to distinguish all four polarization Bell states (near deterministically). At the same time, Nemoto and Munro \cite{Nemoto2004} constructed a near deterministic controlled-NOT gate using several single photon sources and two QND detections. Also, in their Letter, a near deterministic entangler was presented to entangle two separable polarization qubits. The cross-Kerr nonlinearity has
a Hamiltonian of the form \({\mathcal{H}_{QND}} = \hbar \chi
a_s^\dag {a_s}a_p^\dag {a_p}\),  where $a_s^\dag$ and ${a_s}$ ($a_p^\dag$ and ${a_p}$)
represent the creation and annihilation operators of the signal
mode (probe mode) and $\chi$ is the coupling
strength of the nonlinearity. Let us consider the combined system  of a signal mode in a state ${\left| \psi
\right\rangle _s} = {\lambda_1} {\left| 0 \right\rangle _s} + {\lambda_2}{\left|
1 \right\rangle _s}+  {\lambda_3}{\left| 2\right\rangle _s}$ and a  probe beam in a coherent state $|\alpha\rangle_p$.
 When the signal photons interacting with the cross-Kerr media and inducing a phase shift on the probe beam the whole system evolves as
 \({\texttt{e}^{\texttt{i}{\mathcal{H}_{QND}}t/\hbar }}{\left| \psi  \right\rangle _s}{\left| \alpha  \right\rangle _p}
 = {\lambda_1}{\left| 0 \right\rangle _s}{\left| \alpha  \right\rangle _p} + {\lambda_2}{\left| 1 \right\rangle _s}{\left|
 {\alpha {\texttt{e}^{\texttt{i}\theta }}} \right\rangle _p}+{\lambda_3}{\left| 2\right\rangle _s}{\left|
 {\alpha {\texttt{e}^{2\texttt{i}\theta }}} \right\rangle _p}\), where \(\theta  = \chi t\) and $t$ is the
 interaction time. Then, a homodyne measurement connected with a quadrature operator $\hat{x}(\phi)=a_p\texttt{e}^{\texttt{i}\phi}+a_p^\dag\texttt{e}^{-\texttt{i}\phi}$ acts on the probe beam to project the signal mode onto an expected subspace. Here $\phi$ is a real constant. The homodyne measurement is called an $X$ quadrature measurement when $\phi=0$ \cite{Barrett2005, Nemoto2004} (or a $P$ quadrature measurement if $\phi=\frac {\pi}{2}$ \cite{Munro2005}) for a real initial coherent state.

Although there exist various studies for generating the maximally entangled multi-qubit states \cite{Zou03, Duan2003, Leibfried2005, Jin Song2009, Wang2010, Feng2011}, we here restrict our discussion to the situation in quantum optics. In 1997, Zeilinger \emph{et al}. \cite{Zeilinger97} put forward the main idea for creating a three-photon GHZ state from two entangled-photon pairs with linear optics. That is, when a single particle from two independent entangled pairs is detected, under the condition that it is impossible to determine from which pair the single particle comes, the remaining three particles become entangled in a GHZ state. Shortly thereafter, the three-photon GHZ entanglement was observed by Bouwmeester \emph{et al}. \cite{GHZExperiment99} in experiment, and then the observation of highly pure four-photon GHZ entanglement \cite{Pan2001}. Later, Sagi \cite{Sagi2003} presented a scheme for the probabilistic creation of $n$-photons GHZ-type states with also linear optics and single-photon detectors. Based on QND measurement, a near deterministic two-photon entangler \cite{Nemoto2004}, a scheme for generating cluster states including a three-photon GHZ entangler \cite{Louis2007}, and several multiphoton GHZ entangler \cite{Jin2007, Wang2011} were presented. However, there exists a difficulty for realizing the schemes since the conditional phase shifts with opposite signs have been introduced \cite{Kok2008}. The difficulty can be overcome by simply introducing a phase gate independent of the cross-Kerr media, followed by an appropriate measurement \cite{MNS2005, LinHeBR2009, DingYan2012}. On the other hand, to distinguish between all four Bell states with linear optics \cite{Schuck2006, Pavicic2011}, a $50:50$ beam splitter \cite{Braunstein1995} and Hong-Ou-Mandel interferometer \cite{HOM1987} have been introduced. For three-photon GHZ-state, Pan and Zeilinger \cite{Pan1998} firstly proposed a scheme to construct a GHZ-state analyzer based on linear-optics elements. Recently, Qian \emph{et al}. \cite{Qian2005} presented a scheme of three-photon GHZ-state analyzer using two-photon polarization QND parity detectors, in which all of the eight three-photon GHZ states can be near deterministically discriminated.

The outline of this paper is as follows. First we present a scheme of entangler for three-photon GHZ-state based on the weak nonlinearities. Several cross-Kerr nonlinearities and a coherent probe beam are used in the scheme. After an $X$ quadrature measurement and a series of appropriate operations on specified qubits according to classical feed-forward information, the desired three-photon GHZ state can be obtained. Then we extend the three-photon entangler to $n$-photon cases   as well as two-photon entangler that generates  the two-photon maximally entangled states i.e.  Einstein-Podolsky-Rosen (EPR) pairs. In Sec.III, we construct a three-photon GHZ-state analyzer with the cross-Kerr nonlinearities. In the  scheme, eight three-photon GHZ states can be coarsely recognized by implementing the $X$ quadrature measurement on the probe beam and then can be completely distinguished by using linear optical elements and single-photon detectors. Later on, we  generalize the three-photon GHZ-state analyzer to the $n$-photon cases. Furthermore, we also describe an efficient nondestructive Bell-state analyzer.

\section{entangler for multiphoton maximally entangled states }

\begin{figure}
  \includegraphics[width=5in]{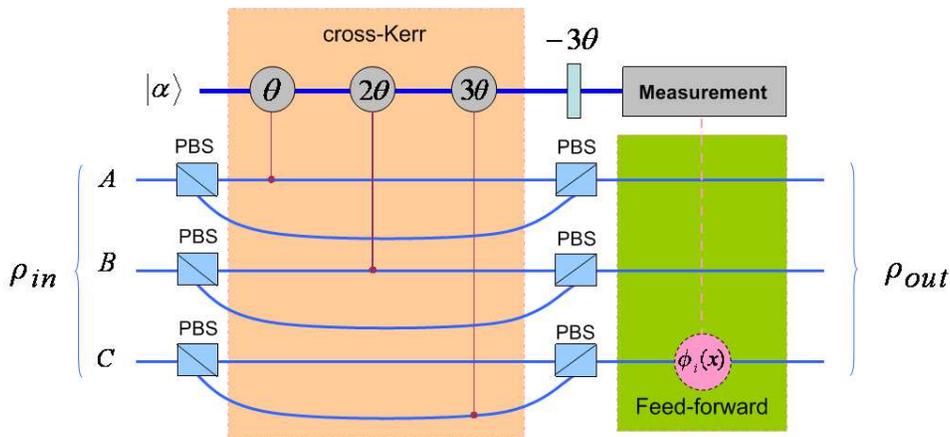}\\
  \caption{(color online). An entangler for three-photon GHZ-state. Consider that three photons entering the entangler from input ports $A$, $B$, and $C$ are prepared in the initial states $\left| {{u _1}} \right\rangle=\left|{{u _2}} \right\rangle=\left|{{u _3}} \right\rangle=\frac{1}{\sqrt{2}}(|H\rangle+|V\rangle)$, respectively. Each polarizing beam splitter (PBS) transmits horizontally polarized photons and reflects vertically polarized ones. Several cross-Kerr nonlinearities, a phase shift gate and an $X$ quadrature measurement are necessary to discriminate between four output states. Each ${\phi _i}\left( x \right) (i=1, 2, 3)$ represents a phase shift on any one of the three qubits via classical feed-forward information.}
  \label{3ghz}
\end{figure}

Now we first  construct a three-photon GHZ-state entangler by using weak nonlinearities and then extend the scheme to the situation of $n$-photons. The setup of our scheme is shown in Fig.\ref{3ghz}. In signal modes, three polarization qubits are initially prepared in the states
  $\left| {{u _1}} \right\rangle  = {a_1}\left| H \right\rangle  + {b_1}\left| V \right\rangle $, $\left| {{u _2}} \right\rangle  = {a_2}\left| H \right\rangle  + {b_2}\left| V \right\rangle $, and $\left| {{u _3}} \right\rangle  = {a_3}\left| H \right\rangle  + {b_3}\left| V \right\rangle$, respectively.
  The total input state for the three signal photons and a probe beam $\left| \alpha \right \rangle$ reads
     $ \left|\psi\right\rangle _{\text{Total}}=
     ({a_1 a_2 a_3} \left| {HHH} \right\rangle  + {b_1 b_2 b_3} \left| {VVV} \right\rangle
         +  {a_1 b_2 b_3} \left| {HVV} \right\rangle  + {b_1 a_2 b_3} \left| {VHV} \right\rangle
           +{a_1 b_2 a_3} \left| {HVH} \right\rangle  + {b_1 a_2 a_3} \left| {VHH} \right\rangle
          + {a_1 a_2 b_3} \left| {HHV} \right\rangle  + {b_1 b_2 a_3} \left| {VVH} \right\rangle)
  \left| \alpha \right \rangle $,
  where $a_i, b_i$ $(i=1, 2, 3)$ are complex coefficients satisfying the normalization requirements respectively. As a matter of fact,  the following discussions are suitable to the three-photon state which is in the most general polarization superposition, i.e.,
  $ \left|\psi\right\rangle=
     \alpha_1 \left| {HHH} \right\rangle  + \alpha_2\left| {VVV} \right\rangle
         +  \alpha_3 \left| {HVV} \right\rangle  + \alpha_4 \left| {VHV} \right\rangle
           +\alpha_5 \left| {HVH} \right\rangle  + \alpha_6\left| {VHH} \right\rangle
          + \alpha_7 \left| {HHV} \right\rangle  + \alpha_8\left| {VVH} \right\rangle$, where $\alpha_i$ ($i=1,2,...,8$) are complex coefficients satisfying $\sum_i|\alpha_i|^2=1$.

\begin{figure}
  \includegraphics[width=3in]{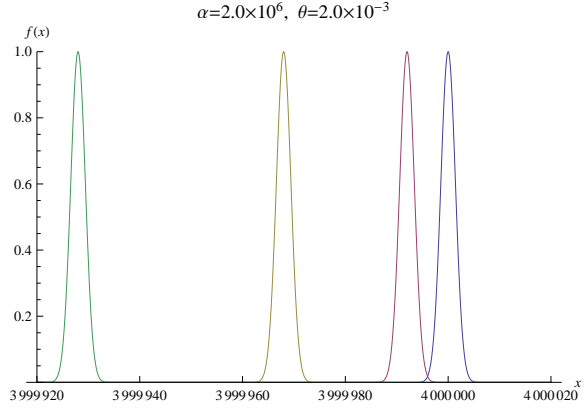}\\
  \caption{(color online). Plot of the Gaussian distributions for the outputs of the $X$ quadrature measurement on the probe beam. $f(x)$ is Gaussian term associated with $x$---the possible measurement value of the observable $X$. Four peaks are, from right to left, associated with the states ${\left|\psi_1\right\rangle}$, ${\left|\psi_2\right\rangle}$, ${\left|\psi_3\right\rangle}$, and ${\left|\psi_4\right\rangle}$, respectively.}
  \label{pd}
\end{figure}

In Fig.\ref{3ghz}, input qubits in signal modes are individually split into two spatial modes on PBSs and one mode of each qubit interacts with a
weak cross-Kerr nonlinearity to pick up a phase shift $\theta$ ($2\theta$, or $3\theta$) on the coherent probe beam. After all  qubits have interacted with the nonlinear media and a further phase shift (independent of the nonlinear media) of $-3\theta$ has been applied to the probe beam, the whole combined system evolves into

$$\left|\psi\right\rangle_\text{ck}  = \left( {a_1 a_2 a_3 \left| {HHH} \right\rangle + b_1 b_2 b_3 \left| {VVV} \right\rangle } \right)\left| \alpha  \right\rangle
  + a_1 b_2 b_3 \left| {HVV} \right\rangle \left| {\alpha \texttt{e}^{\texttt{i}\theta } } \right\rangle
  + b_1 a_2 a_3 \left| {VHH} \right\rangle \left| {\alpha \texttt{e}^{ - \texttt{i}\theta } } \right\rangle $$
  \begin{equation}\label{}
  + b_1 a_2 b_3 \left| {VHV} \right\rangle  \left| {\alpha \texttt{e}^{2\texttt{i}\theta } } \right\rangle
  + a_1 b_2 a_3 \left| {HVH} \right\rangle \left| {\alpha \texttt{e}^{-2\texttt{i}\theta } } \right\rangle
  + a_1 a_2 b_3 \left| {HHV} \right\rangle \left| {\alpha \texttt{e}^{3\texttt{i}\theta } } \right\rangle
  + b_1 b_2 a_3 \left| {VVH} \right\rangle \left| {\alpha \texttt{e}^{ - 3\texttt{i}\theta } } \right\rangle.
  \end{equation}\\
We observe immediately that the components $|HHH\rangle$ and $|VVV\rangle$  pick up no phase shift, while the components $|HVV\rangle$ and $|VHH\rangle$, $|VHV\rangle$ and $|HVH\rangle$, or $|HHV\rangle$ and $|VVH\rangle$ induce opposite sign phase shift $\theta$, $2\theta$, or $3\theta$, respectively. There exist various measurements on the probe beam to realize the entangler conditioning.  We now perform an $X$ quadrature measurement \cite{Barrett2005} -- a straightforward but efficient measurement strategy -- on the probe beam. When one choose the local oscillator phase $\pi/2$ offset from the probe phase, with $|\alpha\rangle$ real, the states $\left| {\alpha \texttt{e}^{ \pm {\texttt{i}\theta }} } \right\rangle$ cannot be distinguished \cite{Nemoto2004}, similar to the states $\left| {\alpha \texttt{e}^{ \pm 2\texttt{i}\theta } } \right\rangle$ and $\left| {\alpha \texttt{e}^{ \pm 3\texttt{i}\theta } } \right\rangle$. After the homodyne measurement on the probe beam the three-photon state yields
$$\left|\psi\right\rangle _X = f\left( {x,\alpha } \right)\left( {a_1 a_2 a_3 \left| {HHH} \right\rangle  + b_1 b_2 b_3 \left| {VVV} \right\rangle } \right)
+ f\left( {x,\alpha \cos\theta } \right)\left( {a_1 b_2 b_3 \texttt{e}^{\texttt{i}\phi _1 \left( x \right)} \left| {HVV} \right\rangle
+ b_1 a_2 a_3 \texttt{e}^{ - \texttt{i}\phi _1 \left( x \right)} \left| {VHH} \right\rangle } \right)$$
$$+ f\left( {x,\alpha \cos2\theta } \right)\left( {b_1 a_2 b_3 \texttt{e}^{\texttt{i}\phi _2 \left( x \right)} \left| {VHV} \right\rangle  + a_1 b_2 a_3 \texttt{e}^{ - \texttt{i}\phi _2 \left( x \right)} \left| {HVH} \right\rangle } \right)$$
  \begin{equation}\label{}
+ f\left( {x,\alpha \cos3\theta } \right)\left( {a_1 a_2 b_3 \texttt{e}^{\texttt{i}\phi _3 \left( x \right)} \left| {HHV} \right\rangle  + b_1 b_2 a_3 \texttt{e}^{ - \texttt{i}\phi _3 \left( x \right)} \left| {VVH} \right\rangle } \right),
 \end{equation}
where $f\left( {x,\beta } \right) = {\left( {2\pi } \right)^{ - 1/4}}{\texttt{e}^{ - \left( {x - 2{\beta}} \right)^2/4}}$ and ${\phi _i}\left( x \right) = \alpha \sin i \theta \left( {x - 2\alpha \cos i \theta } \right)\bmod 2\pi$,  $(i=1, 2, 3)$.
 $f\left( {x,\alpha } \right)$, $f\left( {x,\alpha \cos \theta } \right)$, $f\left( {x,\alpha \cos 2\theta } \right)$ and $f\left( {x,\alpha \cos 3\theta } \right)$ are  respectively the Gaussian curves with the peaks located at $2\alpha$, $2\alpha \cos\theta$, $2\alpha \cos2\theta$ and $2\alpha \cos3\theta$ (depicted in Fig.\ref{pd}). These curves corresponding to probability amplitudes associate with the outputs of the signal photons. ${\phi _1}\left( x \right)$, ${\phi _2}\left( x \right)$ and ${\phi _3}\left( x \right)$ are three phase shift operations corresponding to the value of $x$---the outputs of the $X$ quadrature measurement. Each of these operations performs a conditional phase shift to evolve the three-photon system into a desired state, up to an unobservable global phase factor. The midpoints between two neighboring peaks are ${x_{{m_1}}} = \alpha \left( {\cos 2\theta  + \cos 3\theta } \right)$, ${x_{{m_2}}} = \alpha \left( {\cos \theta  + \cos 2\theta } \right)$ and ${x_{{m_3}}} = \alpha \left( {1 + \cos \theta } \right)$. Two nearby peaks are respectively separated by the distances ${x_{{d_1}}} = 2\alpha \left( {\cos 2\theta  - \cos 3\theta } \right) \sim 5\alpha {\theta ^2}$, ${x_{{d_2}}} = 2\alpha \left( {\cos \theta  - \cos 2\theta } \right) \sim 3\alpha {\theta ^2}$ and ${x_{{d_3}}} = 2\alpha \left( {1  - \cos \theta } \right) \sim \alpha {\theta ^2}$. So the corresponding relationship between the measurement result $x$ and the signal quantum state is
 \begin{equation}\label{x homodyne}
 {\left|\psi\right\rangle _X} \sim \left\{ \begin{gathered}
  {\left|\psi_1\right\rangle}={a_1}{a_2}{a_3}\left| {HHH} \right\rangle  + {b_1}{b_2}{b_3}\left| {VVV} \right\rangle, ~~~~~~~~~~~~~~~~~~~~~~~~~~~~~~~\texttt{for}~~x > {x_{{m_3}}}, \hfill \\
  {\left|\psi_2\right\rangle}={a_1 b_2 b_3 \texttt{e}^{\texttt{i}\phi _1 \left( x \right)} \left| {HVV} \right\rangle
+ b_1 a_2 a_3 \texttt{e}^{ - \texttt{i}\phi _1 \left( x \right)} \left| {VHH} \right\rangle }, ~~~~~~~~~~~~\texttt{for}~~{x_{{m_2}}} < x < {x_{{m_3}}}, \hfill \\
{\left|\psi_3\right\rangle}={b_1 a_2 b_3 \texttt{e}^{\texttt{i}\phi _2 \left( x \right)} \left| {VHV} \right\rangle  + a_1 b_2 a_3 \texttt{e}^{ - \texttt{i}\phi _2 \left( x \right)} \left| {HVH} \right\rangle }, ~~~~~~~~~~~~\texttt{for}~~{x_{{m_1}}} < x < {x_{{m_2}}}, \hfill \\
  {\left|\psi_4\right\rangle}={{a_1}{a_2}{b_3}\texttt{e}^{\texttt{i}{\phi _3}\left( x \right)}}\left| {HHV} \right\rangle  + {{b_1}{b_2}{a_3}\texttt{e}^{ - \texttt{i}{\phi _3}\left( x \right)}}\left| {VVH} \right\rangle, ~~~~~~~~~~~~\texttt{for}~~x < {x_{{m_1}}}. \hfill \\
\end{gathered}  \right.
\end{equation}
 Evidently,  there exist four intervals of the results of the homodyne measurement and each interval connects with an output state of the signal photons. In fact, the error probabilities due to the overlaps between neighboring curves are given by $\varepsilon_i = { \text{erfc}\left( {{x_{d_i}}/2\sqrt 2 } \right)}/2$, where $i=1, 2, 3$. If the distances are large---only small overlaps between these curves---there are very small error probabilities. We consider $\alpha \theta^2  \sim 8$, but in the regime of weak nonlinearities ($\theta \ll 1$ ), as described in the scheme  proposed by Nemoto and Munro \cite{Nemoto2004}. For instance, if $\alpha =2.0\times10^6$ and $\theta=2.0\times10^{-3}$ (realistic techniques),  the maximal value of the error probabilities is  $\varepsilon_{\text{max}} = \varepsilon_3 \sim 3\times10^{-5}$. Thereby, in a sense, the scheme can be realized in a near deterministic manner.

By Eq.(\ref{x homodyne}), we know that  a family of three-photon entangled states can be created, for example, ${\left|\psi_1\right\rangle}={a_1}{a_2}{a_3}\left| {HHH} \right\rangle  + {b_1}{b_2}{b_3}\left| {VVV} \right\rangle$ with the appropriate choice of $a_i$ and $b_i$. Especially, when we choose $a_i=b_i=1/\sqrt 2 $, the output state is one of the maximally entangled states---three-photon GHZ states. Subsequently, because of the equivalence of the eight GHZ states under local operations and classical communication, the present scheme can, of course, generate the specified three-photon GHZ state $|\psi_0\rangle=(|HHH\rangle+|VVV\rangle)/\sqrt{2}$ by means of classical feed-forward information. For instance, after the phase shift ${\phi _i}\left( x \right)$ has been performed on any one of these qubits (the third qubit, for example), a further single qubit operation---NOT gate ($\sigma_x=|H\rangle\langle V|+|V\rangle\langle H|$)---is performed on the specified qubit according to the classical feed-forward based on the result of the homodyne measurement, the specified three-photon GHZ state $|\psi_0\rangle$ is obtained. It should be noted that  for $x > {x_{{m_3}}}$, the resulting state is exactly $|\psi_0\rangle$ and no operation needs to be done. Hence, by using this scheme, one can create the desired three-photon GHZ state
near deterministically.

\begin{figure}
  \includegraphics[width=4in]{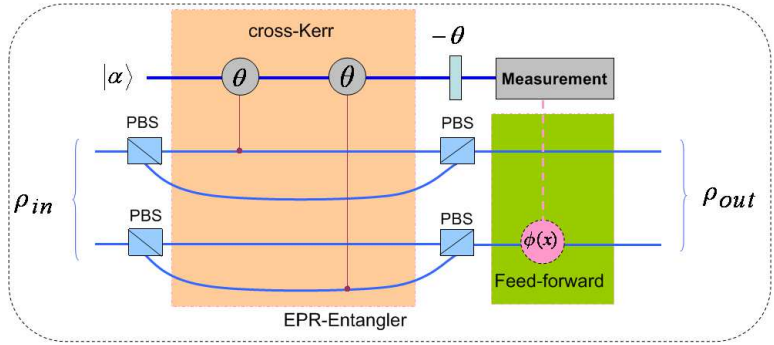}\\
  \caption{(color online) An entangler for two-photon EPR pairs.}
  \label{EPR-entangler}
\end{figure}

\begin{figure}
  \includegraphics[width=5in]{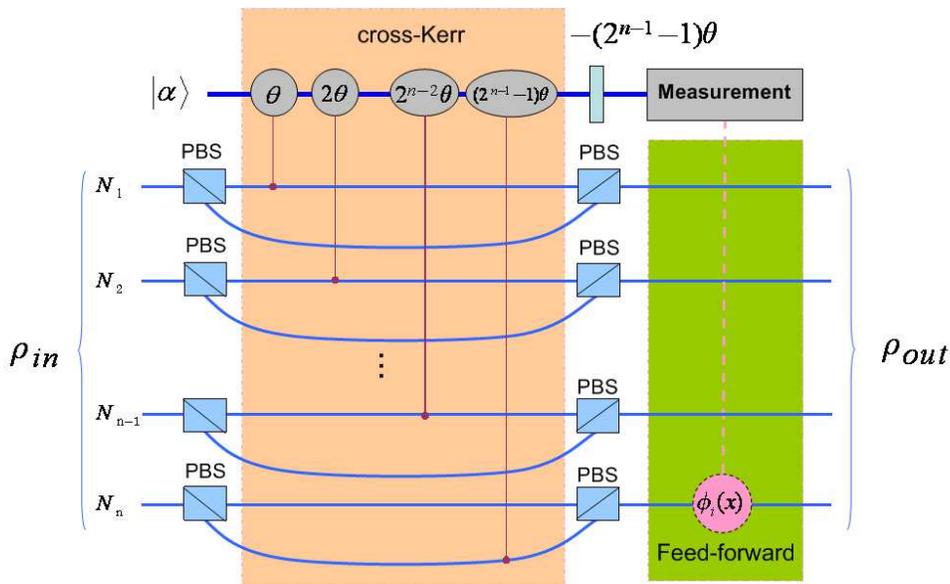}\\
  \caption{(color online). An entangler for $n$-photon GHZ states.}
  \label{nghz}
\end{figure}

The above entangler for three-photon GHZ state can be easily extended to the  $n$-photon cases, where $n\geq2$. For $n=2$, obviously, the scheme becomes a perfect entangler for two-photon EPR pairs shown in Fig.\ref{EPR-entangler} and it is exactly the result of Fig.2 shown in Ref. \cite{MNS2005}  except for the measurement being used. For the sake of simplicity, we suppose  that the initial state of  $n$-photons (see Fig.\ref{nghz})  is
\begin{equation}\label{}
\rho_\text{in}=[(|H\rangle_1+|V\rangle_1)\otimes\cdots\otimes(|H\rangle_n+|V\rangle_n)(_1\langle H|+_1\langle V|)\otimes\cdots\otimes(_n\langle H|+_n\langle V|)]/2^n.
\end{equation}
 After all the qubits have interacted with the cross-Kerr nonlinear media and  a necessary phase gate $-(2^{n-1}-1)\theta$ has been performed, the state of combined system of the qubits and a probe beam becomes
$$|\Phi\rangle_\text{ck}=[(|HH\cdots H\rangle+|VV\cdots V\rangle)|\alpha\rangle+|HV\cdots V\rangle|\alpha \texttt{e}^{\texttt{i}\theta}\rangle+|VH\cdots H\rangle|\alpha \texttt{e}^{-\texttt{i}\theta}\rangle+\cdots$$
\begin{equation}\label{}
+|H\cdots HV\rangle|\alpha \texttt{e}^{(2^{n-1}-1)\texttt{i}\theta}\rangle+|V\cdots VH\rangle|\alpha \texttt{e}^{-(2^{n-1}-1)\texttt{i}\theta}\rangle]/\sqrt{2^n}.
\end{equation}
Similarly, after an $X$ quadrature measurement and subsequent local operations---a phase shift and one or more NOT gates on specified qubits---according to the classical feed-forward, the  state can be converted to the $n$-photon GHZ state $\rho_\text{out}=(|H\rangle^{\otimes n}+|V\rangle^{\otimes n})(^{\otimes n}\langle H|+^{\otimes n}\langle V|)/2$. Notice that with the increase of the number of the photons the intensity of nonlinearities required is larger and larger. Therefore, the number of the photons of a generating GHZ state in the present scheme depends mainly on the improvement of the nonlinearity. On the other hand, the maximal error probability is still equal to $\varepsilon_n \sim 3\times10^{-5}$. So this scheme can also be realized in a near deterministic manner.

There are two main differences between the present scheme and  Jin's scheme \cite{Jin2007}. First, in our scheme it uses  a coherent probe beam   and an $X$-quadrature measurement instead of  a coherent state
superposition  and a $P$-quadrature measurement as in Jin's protocol. Second,  in our scheme there is  no the minus phase shift, which appears in Jin's scheme. As a matter of fact, we  arrive at this goal  by connecting $n-$th photon's vertical polarization with the coherent probe beam such that the  interaction between the  photons  and the cross-Kerr media  only induces positive phase shifts $\theta$, $2\theta$, $\cdots$, $(2^{n-1}-1)\theta$ as shown in Fig.\ref{nghz}.

\section{analyzer for multiphoton maximally entangled states }
\begin{figure}
  \includegraphics[width=3in]{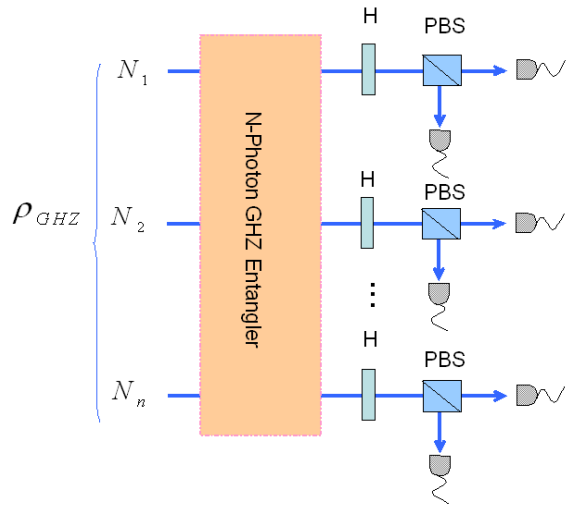}\\
  \caption{(color online). An analyzer for $n$-photon GHZ states. Each $H$ represents a half-wave plate which is used to implement an $H$ gate operation (the angle between its axis and the horizontal direction is $22.5^{\circ}$). The $2n$ single-photon detectors on the right-hand side are used to discriminate between two signs ``$\pm$" in each class.}
  \label{lo-analyzer}
\end{figure}

As described above, an entangler for $n$-photon maximally entangled states has been presented. It is worth noting that the entangler can be utilized to constitute an analyzer to distinguish orthogonal $n$-photon maximally entangled GHz states, as shown in Fig.\ref{lo-analyzer}. For three-photon case,  the eight maximally entangled GHZ states are given by
$
{\left| {\phi _1^ \pm } \right\rangle _{N_1N_2N_3}} = \frac{1}
{{\sqrt 2 }}{\left( {\left| {HHH} \right\rangle  \pm \left| {VVV} \right\rangle } \right)_{N_1N_2N_3}},
{\left| {\phi _2^ \pm } \right\rangle _{N_1N_2N_3}} = \frac{1}
{{\sqrt 2 }}{\left( {\left| {HVV} \right\rangle  \pm \left| {VHH} \right\rangle } \right)_{N_1N_2N_3}},
{\left| {\phi _3^ \pm } \right\rangle _{N_1N_2N_3}} = \frac{1}
{{\sqrt 2 }}{\left( {\left| {HVH} \right\rangle  \pm \left| {VHV} \right\rangle } \right)_{N_1N_2N_3}},
$ and
$
{\left| {\phi _4^ \pm } \right\rangle _{N_1N_2N_3}} = \frac{1}
{{\sqrt 2 }}{\left( {\left| {HHV} \right\rangle  \pm \left| {VVH} \right\rangle } \right)_{N_1N_2N_3}}.
$
We suppose that an input state of the above entangler, as shown in Fig.\ref{lo-analyzer} for $n=3$, is one of the states ${\left| {\phi _i^ \pm } \right\rangle _{N_1N_2N_3}}$ ($i=1, 2, 3, 4$), and each photon enters modes $N_1$, $N_2$, and $N_3$, respectively. By the above discussion, there would be four cases. That is, for $x>x_{m_3}$, the input state must be one of the two states $\left| {\phi _1^ \pm } \right\rangle _{N_1N_2N_3}$; for ${x_{{m_2}}} < x < {x_{{m_3}}}$, the input state belongs to one of the two states $\left| {\phi _2^ \pm } \right\rangle _{N_1N_2N_3}$; for ${x_{{m_1}}} < x < {x_{{m_2}}}$, the input state must be one of the two states $\left| {\phi _3^ \pm } \right\rangle _{N_1N_2N_3}$; and for $x<x_{m_1}$, the input state is one of the two states $\left| {\phi _4^ \pm } \right\rangle _{N_1N_2N_3}$. Let us  classify the eight GHZ states into four classes,  ${\left| {\phi _1^ \pm } \right\rangle _{N_1N_2N_3}}$, ${\left| {\phi _2^ \pm } \right\rangle _{N_1N_2N_3}}$, ${\left| {\phi _3^ \pm } \right\rangle _{N_1N_2N_3}}$, and ${\left| {\phi _4^ \pm } \right\rangle _{N_1N_2N_3}}$. Apparently, it enables us to recognize which class the initial state belongs to via the result of $X$ quadrature measurement on the probe beam.

  Furthermore, the signs ``$\pm$" of each class can be  discriminated by using linear optical elements and single-photon detectors \cite{Pan1998, Qian2005}. Without loss of generality, let us consider the states ${\left| {\phi _1^ \pm } \right\rangle _{N_1N_2N_3}}$. When each of three photons respectively passes through a half-wave plate,    the state ${\left| {\phi _1^ + } \right\rangle _{N_1N_2N_3}}$ becomes $  \frac {1}{2}(|HHH\rangle+|HVV\rangle+|VHV\rangle+|VVH\rangle)_{N_1N_2N_3}$, and ${\left| {\phi _1^ - } \right\rangle _{N_1N_2N_3}}$ evolves to $ \frac {1}{2}(|HHV\rangle+|HVH\rangle+|VHH\rangle+|VVV\rangle)_{N_1N_2N_3}$ (see Eqs. (11) and (12) described in Ref. \cite{Pan1998} ). After being transmitted or reflected from the polarizing beam splitters, the photons will enter the  single-photon detectors, as shown in Fig.\ref{lo-analyzer}. According to  the measurement  results of the single-photon detectors, we can determine  whether the initial state is  ${\left| {\phi _1^ +} \right\rangle _{N_1N_2N_3}}$ or the  state ${\left| {\phi _1^ -} \right\rangle _{N_1N_2N_3}}$.

The above scheme of three-photon GHZ state analyzer can also be extended to $n$-photon GHZ state analyzer. For $2^n$ orthogonal $n$-photon GHZ states, we can first classify the input states into $2^{n-1}$ classes using an $n$-photon GHZ entangler which can not discriminate between the signs ``$\pm$" in each classes. In order to further distinguish one state from the other, a series of linear optical elements---half-wave plates---and single-photon detectors are designed as shown in Fig.\ref{lo-analyzer}.

\begin{figure}
  \includegraphics[width=3in]{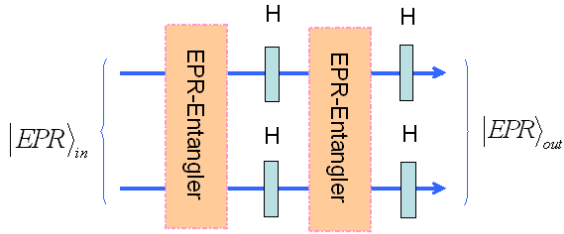}\\
  \caption{(color online) The schematic diagram of two-step nondestructive Bell-state analyzer with the cross-Kerr nonlinearities.}
  \label{EPR-analyzer}
\end{figure}

For nondestructive Bell-state detection, as an example, we describe a scheme of two-step nondestructive Bell-state analyzer with the cross-Kerr nonlinearities shown in Fig.\ref{EPR-analyzer}. First, we can conclude that whether the initial state belongs to the states $(|00\rangle\pm|11\rangle)/\sqrt{2}$ connecting with the result $x>x_m$ or the states $(|01\rangle\pm|10\rangle)/\sqrt{2}$ associating with $x<x_m$ using a two-photon entangler shown in Fig.\ref{EPR-entangler}, where $|0\rangle$ and $|1\rangle$ denote horizontal and vertical polarization of the photons, respectively, and  $x_m$ is the midpoint between two peaks of the Gaussian curves. We then apply an $H$ gate on each photon and a further two-photon entangler to distinguish one state from another according to the second result of $x$. At last, a further $H$ gate is performed on each photon to reconvert the output state into the initial one. An efficient nondestructive Bell-state detection, therefore, can be realized based on two-step two-photon entangler, the details  are shown in Table.\ref{TAB1}.

\renewcommand\arraystretch{1.5}
\begin{table}[htbp]
 \caption{The results of the $X$ quadrature measurements and the corresponding states for the scheme of two-step nondestructive Bell-state analyzer with the cross-Kerr nonlinearities.}\label{TAB1}

 \tabcolsep0.15in
 \doublerulesep2pt
 \begin{tabular}{ccccc}
  \hline\hline
 \ inputs   & $x$ (first)   & $H$ &$x$ (second) & $H$ (outputs)\\
 \hline

\ $(|00\rangle+|11\rangle)/\sqrt{2}$& $x>x_m$ & $(|00\rangle+|11\rangle)/\sqrt{2}$  &   $x>x_m$  &$(|00\rangle+|11\rangle)/\sqrt{2}$\\
\ $(|00\rangle-|11\rangle)/\sqrt{2}$& $x>x_m$ & $(|01\rangle+|10\rangle)/\sqrt{2}$  &   $x<x_m$  &$(|00\rangle-|11\rangle)/\sqrt{2}$\\
\ $(|01\rangle+|10\rangle)/\sqrt{2}$& $x<x_m$ & $(|00\rangle-|11\rangle)/\sqrt{2}$  &   $x>x_m$  &$(|01\rangle+|10\rangle)/\sqrt{2}$\\
\ $(|01\rangle-|10\rangle)/\sqrt{2}$& $x<x_m$ & $(|10\rangle-|01\rangle)/\sqrt{2}$  &   $x<x_m$  &$(|01\rangle-|10\rangle)/\sqrt{2}$\\
 \hline
 \hline
 \end{tabular}

\end{table}

\section{discussion and summary}

We have described a new scheme of an entangler for multiphoton maximally entangled states---$n$-photon GHZ states and Bell states, where several linear optical elements and a quantum nondemolition measurement are adopted. In the scheme, because only a product state of $n$ photons is supplied instead of preparing a large number of two-photon entangled states, thereby, one can generate the desired $n$-photon GHZ states directly instead of waiting the supply of two-photon entangled states. Also, we have proposed an analyzer for $n$-photon GHZ-state based on the weak nonlinearities, by which one of the $2^n$ maximally entangled GHZ states can be easily classified into $2^{n-1}$ coarse grained classes. Using several linear optical elements and single-photon detectors, at last, all of the $2^n$ $n$-photon GHZ states can be discriminated. Especially, our new scheme is available for analyzing four Bell-state efficiently. In a word, we have proposed two schemes for generating and detecting multiphoton maximally entangled states with the weak nonlinearities and they can be realized nearly deterministically.

This work was supported by the National Natural Science Foundation
of China under Grant No: 10971247, Hebei Natural Science Foundation
of China under Grant Nos: A2012205013, A2010000344, the Fundamental Research Funds for the Central Universities of Ministry of Education of China under Grant No:2011B025.

\end{document}